\documentclass{ws-procs975x65}

\begin{document}

\title{A REVIEW OF THE $1/N$ EXPANSION IN RANDOM TENSOR MODELS
}

\author{R. GURAU $^*$}

\address{Perimeter Institute\\
31 Caroline St. N, Waterloo, ON, N2L 2Y5, Canada\\
$^*$E-mail: rgurau@perimeterinstitute.ca\\
http://www.perimeterinstitute.ca}

\begin{abstract}
Matrix models are a highly successful framework for the analytic study of
random two dimensional surfaces with applications to quantum gravity in
two dimensions, string theory, conformal field theory, statistical physics in
random geometry, etc. Their success relies crucially on the so called $1/N$ expansion 
introduced by 't Hooft. 

In higher dimensions matrix
models generalize to tensor models. In the absence of a viable $1/N$ expansion
tensor models have for a long time been less successful in providing
an analytically controlled theory of random higher dimensional topological
spaces. This situation has drastically changed recently. Models for a generic
complex tensor have been shown to admit a $1/N$ expansion dominated by
graphs of spherical topology in arbitrary dimensions and to undergo a phase
transition to a continuum theory. 
\end{abstract}


\bodymatter

\section{Introduction}\label{aba:sec1}

Random matrices were introduced by Wishart \cite{wishart} for the statistical analysis of large samples
and first used in physics by Wigner \cite{wigner} for the study of the spectra of heavy atoms.
An invariant matrix ensemble is a probability distribution for a random $N\times N$ matrix $M$ 
with probability density\footnote{Where $M^{\dagger}$ denotes the hermitian conjugate} 
$\frac{1}{Z} e^{-N \text{Tr} S(MM^{\dagger})}$ for some polynomial $S$
(or $ \frac{1}{Z} e^{-N \text{Tr} S(M)}$ for hermitian matrices $M=M^{\dagger}$).

The moments and partition function of such a probability distribution can be evaluated in perturbations
as sums over ribbon Feynman graphs. The weights of the graphs are fixed by the Feynman rules. 
In his seminal work \cite{'tHooft:1973jz} 't Hooft realized that the size of the matrix $N$ endows an
invariant matrix ensemble with a small parameter, $1/N$, and the perturbative expansion can be reorganized as 
a series in $1/N$ indexed by the genus. 
At leading order in the large $N$ 
limit only planar graphs \cite{Brezin:1977sv} contribute\footnote{Subsequent terms in the $1/N$ 
series can be accessed trough double scaling limits \cite{double,double1,double2}}. 
As the planar graphs form a summable family, invariant matrix models undergo a phase transition to a continuum
theory of random infinitely refined surfaces when the coupling constant 
is tunned to some critical value\cite{Kazakov:1985ds,mm}. 
Matrix models provide the framework for the analytic study of two-dimensional random geometries coupled with 
conformal matter \cite{Kazakov:1986hu, Boulatov:1986sb, Brezin:1989db,Kazakovmulticrit,
Ambjorn:1990ji,Fukuma:1990jw,Makeenko:1991ry,Dijkgraaf:1990rs,Di Francesco:1993nw} and trough the KPZ correspondence 
\cite{Knizhnik:1988ak, david2, DK, Dup} they relate to conformal field theory on fixed geometries.

The success of matrix models inspired their generalization in higher dimensions to random tensor 
models\cite{ambj3dqg,sasa1,mmgravity,sasab,sasac,Oriti:2011jm,Baratin:2011aa} 
in the hope to gain insights into conformal field theory, statistical models in random geometry and 
quantum gravity in three and four dimensions.
In spite of these initial high hopes tensor models have, until recently, failed 
to provide an analytically controlled theory of random geometries: no progress could be 
made because for a long time no generalization of the $1/N$ expansion to tensors has been found. 

The situation has drastically changed with the discovery of the
colored\cite{color,PolyColor,lost} random tensor models. 
The perturbative series of the colored models supports a $1/N$ expansion\cite{Gur3,GurRiv,Gur4} indexed by the {\it degree}, 
a positive integer which plays in higher dimensions the role of the genus, but is {\it not} a topological 
invariant. Leading order graphs, called {\it melonic\cite{Bonzom:2011zz}},
triangulate the $D$-dimensional sphere in any dimension\cite{Gur3,GurRiv,Gur4}
and are a summable family\cite{Bonzom:2011zz}. Like 
their two dimensional counterparts, tensor models undergo a phase transition to a theory of
continuous infinitely refined random spaces when tunning to criticality.
Colored random tensors \cite{coloredreview} give the first analytically accessible theory of random geometries in 
higher dimensions\cite{sefu2,Ryan:2011qm,Carrozza:2011jn,Carrozza:2012kt,IsingD,EDT,doubletens,Bonzom:2012sz,Bonzom:2012qx}.

The results obtained for the colored models extend to all invariant models for a complex
tensor\cite{Bonzom:2012hw}. The colors arise naturally as a canonical bookkeeping 
device tracking the indices of the tensor and provide the key to the  $1/N$ expansion. We 
present in this paper an overview of these results.
The ensuing theory of random tensors generalizing invariant matrix models to higher dimensions is 
universal\cite{Gurau:2011kk,Gurau:2011tj,Gurau:2012ix,Bonzom:2012fu}. Tensor models have 
been generalized to tensor field theories, \cite{BenGeloun:2011rc,BenGeloun:2012pu,
BenGeloun:2012yk,Carrozza:2012uv,Geloun:2012bz,Rivasseau:2011hm} which have been shown to be 
(super) renormalizable and, at least in two instances \cite{BenGeloun:2012pu,BenGeloun:2012yk}, asymptotically free. 

\section{Tensor Models}

Invariant tensor models for a random tensor of rank $D$ are, loosely speaking, probability measures of the form 
\begin{equation} 
 d\nu= \frac{1}{Z } e^{  - N^{D-1} S (T,\bar  T)  } \; \Bigl( \prod dT_{n_1 \dots n_D} 
 d\bar T_{ n_1 \dots n_D } \Bigr)\;,
\end{equation}
where the ``action'' $S (T,\bar  T) $ is some invariant polynomial.

Let $T$ be a covariant complex tensor of rank $D$ transforming under the external tensor product of $D$ fundamental 
representations of the unitary group $\otimes_{i=1}^D U(N_i)$, for some fixed dimensions $N_1,\dots N_D$.
The tensor $T$ can be seen as a collection of $\prod_{i=1}^{D} N_i$ complex 
numbers $T_{n_1\dots n_D}, \; (n_i = 1,\dots N_i )$ transforming as
\begin{eqnarray}
&& T'_{a_1\dotsc a_D} = \sum_{n_1,\dotsc,n_D} U_{a_1n_1}\dotsm V_{a_Dn_D}\ T_{n_1\dotsc n_D}  \crcr
&& \bar T'_{  a_1\dots  a_D} = \sum_{n_1,\dotsc,n_D} \bar U_{  a_D n_D}\dotsm \bar V_{a_1 n_1}\ \bar T_{n_1\dots n_D}  \; .
\end{eqnarray} 
Each unitary group $U(N_i)$ acts independently on its corresponding index. In particular, 
as the dimensions $N_1, \dots N_D$ might very well be different,  $T$ has {\it no symmetry properties}
under permutation of its indices. For simplicity we restrict to $N_i =N$, for all $i$.
The complex conjugate tensor $\bar T_{  n_1 \dots n_D }$ is a contravariant tensor of rank $D$. 
We will denote $\bar n_i$ the indices of $\bar T$ and $\vec n$ the $D$-uple of integers $(n_1, \dotsc, n_D)$.
We call the position of an index its color: $n_1$ has color $1$, $n_2$ has color $2$ and so on. 

\subsection{Invariants}

By the fundamental theorem of classical invariants of the unitary group any invariant function of the tensor $T$ 
can be expressed as a linear combination of {\it trance invariants}\cite{collins}
built by contracting pairs of covariant and contravariant indices 
in a product of tensor entries
\begin{eqnarray}
\text{Tr} (T,\bar T) = \sum_{n_1 \dots} \prod \delta_{n_1,\bar n_1} \;  T_{n_1\dotsc} \dots \bar T_{\bar n_1 \dots } \; ,
\end{eqnarray} 
where all the indices are saturated. Because the unitary group acts independently on each index, 
the contractions must preserve the color: the first index $n_1$ of a $T$ must always contract 
with the first index $\bar n_1$ on some $\bar T$, the second index $n_2$ of $T$ with the second index $\bar n_2$ on some $\bar T$ and so on.

\begin{figure}[t]
\begin{center}
 \includegraphics[width=6cm]{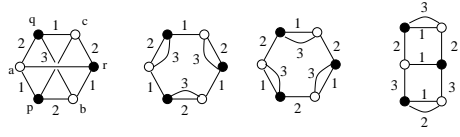}  
\caption{Some trace invariants for $D=3$.}
\label{fig:tensobs}
\end{center}
\end{figure}

The trace invariants can be represented as (bipartite, closed, $D$-colored) graphs. To draw the graph associated 
to a trace invariant we represent every $T_{\dots n_i \dots}$ by a white vertex $v$ and every $\bar T_{\dots \bar n_i \dots}$ by a black vertex $\bar v$.
The contraction of two indices $\delta_{ n_i \bar n_i}$ is represented by a 
line $l^i = (v,\bar v)$ connecting the two corresponding vertices. The lines inherit the color $i$ 
and always connect a black and a white vertex. Some examples of trace invariants for rank 3 tensors are 
represented in figure \ref{fig:tensobs}. For instance the leftmost graph corresponds to the invariant
\begin{eqnarray}
        && \sum \delta_{a_1p_1} \delta_{a_2q_2} \delta_{a_3r_3} \quad   
                    \delta_{b_1r_1} \delta_{b_2p_2} \delta_{b_3q_3} \quad  
                   \delta_{c_1q_1} \delta_{c_2r_2} \delta_{c_3p_3}  \crcr
                && \qquad T_{a_1a_2a_3}  T_{b_1b_2b_3} T_{c_1c_2c_3}  
           \bar T_{p_1 p_2p_3} \bar T_{q_1 q_2q_3} \bar T_{r_1 r_2r_3} \; ,
\end{eqnarray}
where $T_{a_1a_2a_3}  $ is represented by the vertex $a$ in the drawing and so on.
The trace invariant associated to a graph ${\cal B}$ is 
\begin{eqnarray}
 \text{Tr}_{{\cal B}}(T,\bar T ) = \sum   \delta^{{\cal B}}_{\{\vec{n}^v, \vec{\bar{n}}^{\bar v}\}}  \; 
\prod_{v,\bar v  } T_{\vec n^v} \bar T_{\vec {\bar n }^{\bar v} } \;, \quad 
  \text{with} \quad \delta^{{\cal B}}_{\{\vec{n}^v, \bar{\vec{n}}^{\bar v}\}} 
= \prod_{i=1}^D \prod_{l^i = (v,\bar v) } \delta_{n_i^v \bar n_i^{\bar v}} \; ,
\end{eqnarray}
where $v$ (resp. $\bar v$) run over all the white (resp. black) vertices of ${\cal B}$, $l^i$ runs over the lines of color $i$ of ${\cal B}$
and all the indices are summed.
There exists a unique graph with two vertices (all its $D$ lines connect the two vertices). We call it the $D$-dipole and
denote it ${\cal B}_1$. The associated invariant is the unique quadratic invariant
\begin{eqnarray}\label{eq:gaussian}
 \text{Tr}_{{\cal B}_1} (T , \bar T ) = \sum_{\vec n,\vec{\bar {n}}}\, T_{\vec n}\, \bar T_{\vec {\bar n} }
 \ \Bigl[\prod_{i=1}^D \delta_{n_i \bar n_i}\Bigr] \; .
\end{eqnarray} 
We consider in the sequel the most general ``single trace'' tensor model, that is the action is a sum over invariants corresponding
to connected graphs ${\cal B}$
\begin{eqnarray} \label{eq:actiongen}
 S(T,\bar T) =  \, \text{Tr}_{{\cal B}_1} (T , \bar T ) + \sum_{{\cal B} } t_{{\cal B}}\, 
N^{-\frac{2}{(D-2)!} \omega({\cal B})} \,
\text{Tr}_{{\cal B}}(T,\bar T)\;,
\end{eqnarray}
with $t_{\cal B}$ some coupling constants, and we singled out the quadratic part corresponding to ${\cal B}_1$. 
In equation \eqref{eq:actiongen} we have added a scaling in $N^{  -\frac{2}{(D-2)!}\omega( {\cal B}) } $ for every invariant 
which we will fix later. 

The partition function of an invariant tensor model writes then
\begin{equation} 
  Z(t_{\cal B}) = \int \Bigl( \prod_{\vec n =  \vec {\bar n} } dT_{\vec n} d\bar T_{\vec { \bar n} } \Bigr)\;
  e^{ -N^{D-1} S (T,\bar  T)  } \;,
\end{equation}
where $  \vec n =  \vec {\bar n}$ runs over all the $D$-uples of integers $(n_1,\dots n_D), \; n_i = 1\dots N$. The scaling 
$N^{D-1}$ is canonical: it is the only scaling which leads to a well defined large $N$ limit as it can be seen already for 
the Gaussian distribution\cite{Gurau:2011kk}. The observables are (again) the trace invariants 
represented by $D$-colored graphs. 

\begin{figure}[t]
\begin{center}
 \includegraphics[width=3cm]{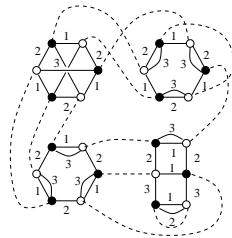}  
\caption{A Feynman graph.}
\label{fig:tensobsgraph}
\end{center}
\end{figure}

The partition function is evaluated 
by Taylor expanding with respect to $t_{{\cal B}}$ and evaluating the Gaussian integral in terms of Wick contractions. 
This leads to a representation in Feynman graphs. The Feynman graphs are made of {\it effective vertices} 
coming from the invariants $ \text{Tr}_{{\cal B}}(T,\bar T) $ (which are graphs ${\cal B}$ with colors $1,\dotsc, D$)
connected by effective \emph{propagators} (Wick contractions, pairings of $T$'s and $\bar T$'s). 
A Wick contraction of two tensor entries $T_{a_1\dotsc a_D}$ and $\bar T_{\bar p_1 \dotsc \bar p_D}$ 
with the quadratic part \eqref{eq:gaussian} consists in replacing them by 
$ \frac{ 1}{N^{D-1} } \prod_{i=1}^D \delta_{a_i \bar p_i} $. We represent the Wick contractions 
by dashed lines to which we assign a new color, $0$. The Feynman graphs (henceforth denoted $ {\cal G} $) are then $D+1$ colored graphs,
see figure \ref{fig:tensobsgraph}.

We denote ${\cal B}_{(\rho)}, \; \rho = 1, \dots |\rho| $ the effective vertices (subgraphs with colors $1, \dots D$)
of a Feynman graph ${\cal G} $. The free energy is a sum over closed, connected $(D+1)$-colored 
graphs
\begin{equation}
F(t_{{\cal B}}) = -\ln Z( t_{{\cal B}})  = \sum_{{\cal G} } \frac{ (-1)^{ |\rho| }}{s({\cal G})}
\, \Bigl( \prod_{\rho=1}^{|\rho|} t_{{\cal B}_{(\rho)}} \Big) \; A({\cal G}) \;,
\end{equation}
where $s({\cal G})$ is a symmetry factor and $A({\cal G})  $ is the amplitude of ${\cal G}$
\begin{eqnarray}\label{eq:ampli}
A({\cal G}) = \sum_{\{\vec{n}^v,\bar{\vec{n}}^{\bar v}\}}\, \Bigl[
\prod_{\rho}  N^{D-1 -\frac{2}{(D-2)!} \omega({\cal B}_{(\rho)} )   } \delta^{{\cal B}_{(\rho)}}_{\{\vec{n}^v,\bar{\vec{n}}^{\bar v}\}}\Bigr] 
 \Bigl[\prod_{l^0=(v,\bar v) } \frac{1}{  N^{D-1}} \prod_{i} \delta_{n_i^v,\bar n_{i}^{\bar v}} \Bigr]\; .
\end{eqnarray}

\subsection{Colored Graphs and topological spaces}

The ribbon graphs of matrix models represent triangulated surfaces. Similarly the colored graphs of tensor models represent triangulated
spaces. This is encoded in the following theorem\cite{lost}.

\begin{theorem}\label{th:pseudo}
  Any closed connected $D+1$ colored graph is a $D$ dimensional normal simplicial pseudo manifold.
\end{theorem}

Loosely speaking a pseudomanifold is a generalization of a manifolds having a finite number of conical singularities. 
One can visualize this pseudomanifold by gluing simplices. We restrict our discussion to the case $D=3$. 
The $D+1$ colored graphs have lines of four colors $0$, $1$, $2$ and $3$. 

\def\figsubcap#1{\par\noindent\centering\footnotesize(#1)}
\begin{figure}[ht]%
\begin{center}
  \parbox{1.5cm}{\epsfig{figure=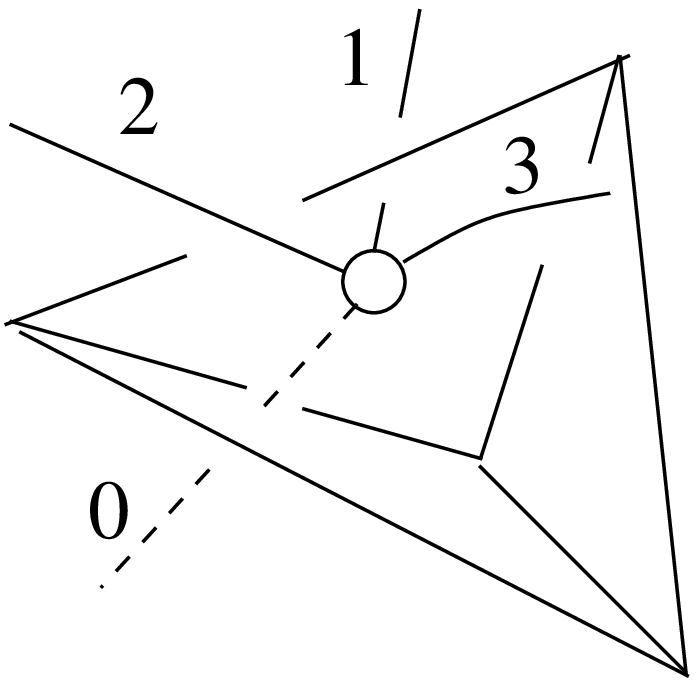,width=1.2cm}\figsubcap{a}}
  \hspace*{8pt}
  \parbox{2.1cm}{\epsfig{figure=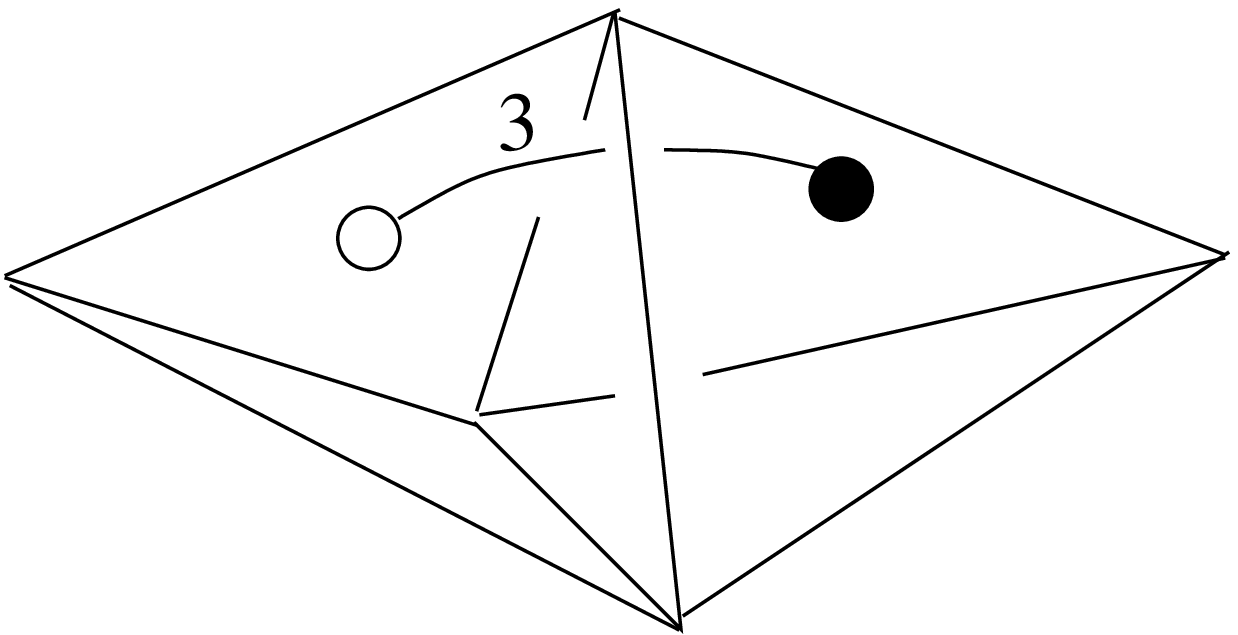,width=2cm}\figsubcap{b}}
   \hspace*{8pt}
  \parbox{2.1cm}{\epsfig{figure=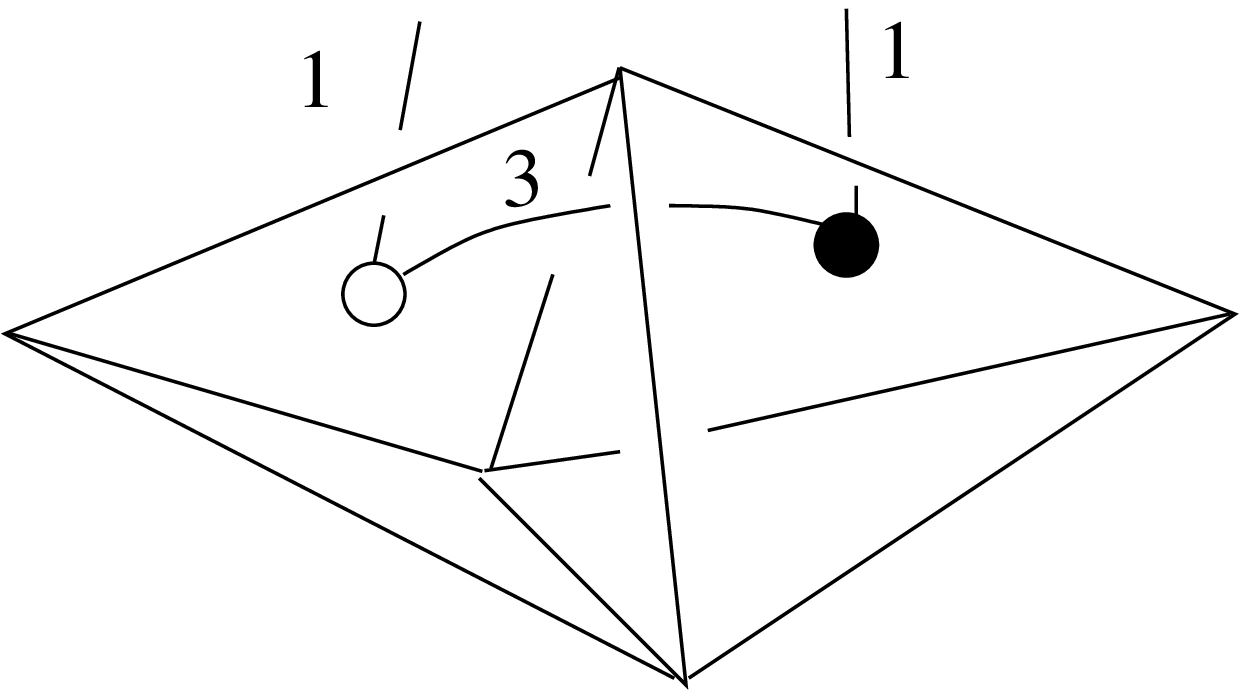,width=2cm}\figsubcap{c}}
  \caption{Gluing of tetrahedra associated to a graph. (a) Vertex. (b) Line. (c) Face.}%
  \label{fig1.2}
\end{center}
\end{figure}

We associate a positive (resp. negative) oriented tetrahedron to every four valent white (resp. black) vertex. 
The triangles bounding the tetrahedron are dual to the lines emanating from the vertex (see figure \ref{fig1.2}(a))
and inherit their color. Thus a tetrahedron is bounded by four triangles of colors $0$, $1$, $2$
and $3$. The coloring of the triangles induces colorings on all the elements of the tetrahedron: the edge of the tetrahedron
common to the triangles $1$ and $2$ is colored by the couple of colors $12$, and so on. Similarly the apex of the tetrahedron 
common to the triangles $1$, $2$ 
and $3$ is colored with the triple of colors $123$. A line in the graph represents the {\it unique} gluing of two tetrahedra
which respects {\it all} the  colorings. Thus (see figure \ref{fig1.2}(b)) we glue the triangle of color $3$ of one tetrahedron on 
the triangle of color $3$ of the other such that the edge $13$ ($23$ resp. $03$) is glued on the 
edge $13$ ($23$ resp. $03$), and the apex $123$ ($023$ resp. $013$) is glued on the apex
$123$ ($023$ resp. $013$). The full cellular structure of the resulting gluing of tetrahedra is encoded in the colors. For 
instance an edge $13$ is represented by a subgraph with colors $13$ in the graph ${\cal G}$ (see figure \ref{fig1.2}(c)). We
call the subgraphs with two colors of ${\cal G}$ its {\it faces}.

Two remarks are in order. First a classical result by Pezzana\cite{Pezzana} ensures that the perturbative expansion 
of tensor models generates all possible manifolds
\begin{theorem}[Pezzana's Existence Theorem]
 Any piecewise linear $D$ dimensional manifold admits a representation as an edge colored graph.
\end{theorem}
In fact for every manifold one can build an infinity of graphs representing it.

\begin{figure}[t]
\begin{center}
 \includegraphics[width=3cm]{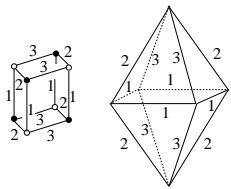}  
\caption{A trace invariant and its associated surface.}
\label{fig:polytope}
\end{center}
\end{figure}

Second, one can ask what is the topological interpretation of an initial trace invariants. As an invariant is a graph with $3$ colors (see figure 
\ref{fig:polytope}), it represents a surface. Adding the lines of color $0$ comes to taking the topological cone over this surface (and if the 
surface has non zero genus it leads to a conical singularity). The 
invariant represents a ``chunk'' of space, and the $3+1$ colored graph can be seem as the gluing of such chunks together. A chunk can alternatively
be seen by associating tetrahedra to the vertices of the invariant (decorated by external halflines of color $0$) and observing that the invariant
represents the gluing of these tetrahedra along triangles of colors $1$ $2$ and $3$, but not along the triangles of color $0$. As in every tetrahedron
the triangle of color zero is opposed to an unique vertex, this chunk is a gluing of tetrahedra (along triangles) around a vertex\cite{Bonzom:2012hw}.

\subsection{The $1/N$ expansion}

We start by a technical prelude. The graphs of matrix models are ribbon graphs made of vertices lines and faces. 
Consider a (closed connected) ribbon graph with an even number of trivalent vertices, $V=2p$, thus having $L = 3p$ lines.
The number of faces of the graph can be expressed as a function of only the number of vertices and the {\it genus} $g$ of the graph 
\begin{equation}
 F = p + 2 - 2g \; .
\end{equation}

A similar relation holds in higher dimensions for $D$ colored graphs. 
To every graph $ {\cal B} $ with $D$ colors we associate a {\it  non negative} integer  
$\omega({\cal B})$ (which we call its {\it degree}) such that the number of faces (subgraph with two colors) of 
${\cal B}$ writes\cite{GurRiv,Gur4}
\begin{equation}
   F  = \frac{(D-1)(D-2)}{2} p + (D-1) - \frac{2}{(D-2)!} \omega({\cal B})\; ,
\end{equation}
with $p$ is the half number of vertices of ${\cal B}$. Naturally a similar relation holds for graphs with
$D+1$ colors by shifting $D$ to $D+1$.

\begin{figure}[ht]%
\begin{center}
  \parbox{1.5cm}{\epsfig{figure=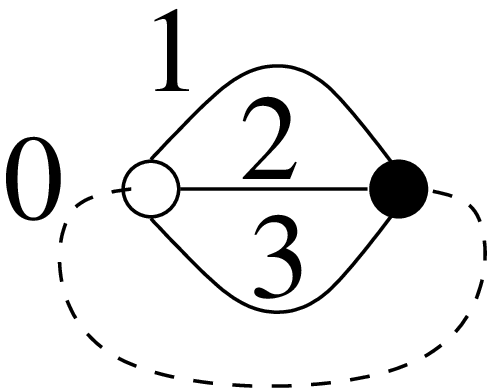,width=1.4cm}\figsubcap{a}}
  \hspace*{8pt}
  \parbox{2cm}{\epsfig{figure=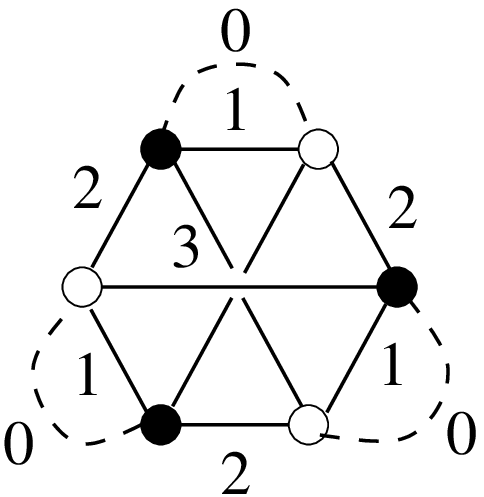,width=1.5cm}\figsubcap{b}}
   \hspace*{8pt}
  \parbox{3cm}{\epsfig{figure=tensobsgraph.eps,width=2.5cm}\figsubcap{c}}
  \caption{3+1 colored graphs of degree (a) $\omega({\cal G})=0$. (b) $\omega({\cal G})=4$. (c) $\omega({\cal G})=10$.}%
  \label{fig:degree}
\end{center}
\end{figure}

The crucial property of the degree is that, like the genus, it is non negative $ \omega({\cal B}) \ge 0 $. 
It is an intrinsic integer number one can compute starting from the graph. However, contrary to the genus, the degree is {\it not} 
a topological invariant but it mixes information about the topology and cellular 
structure\cite{Bonzom:2011br}. The idea is that when counting faces one can identify ribbon graphs ${\cal J}$ 
(called jackets\cite{Geloun:2010nw,Gur3,GurRiv,Gur4,Ryan:2011qm})
embedded in the colored graph. One can separately count the number of faces of each jacket in terms of its 
genus  $g({\cal J})$. Summing over all jackets one gets a counting of the total number of faces of the colored 
graph in terms of the sum of these genera $\omega({\cal B}) \equiv \sum_{{\cal J}} g({\cal J})$ which is the degree of ${\cal B}$.
Some examples (for graphs with $3+1$ colors) are presented in figure \ref{fig:degree}.

We now compute the amplitude of a graph. We fix the scaling of the invariants in the action eq. \eqref{eq:actiongen},  
$\omega({\cal B})$, to be exactly there degree and evaluate eq. \eqref{eq:ampli}. The non trivial part comes from counting the
number of independent sums. Recall that each solid line of colors $1,\dots D$ represents the identification of one index, while the dashed lines
of color $0$ represent the identifications of $D$ indices. It follows that an index, say $n_1$, is identified once along a solid line of color $1$, then
along a dashed line $0$, then along a solid line $1$, the along a dashed line, and so on until the cycle of colors $0$ and $1$ closes. We thus
get a free sum over an index, hence a factor $N$, for every cycle (i.e. face) with colors $0i$.
The total number of faces expresses in terms of the degree of ${\cal G}$ and a short computation yields\cite{Gurau:2011tj}

\begin{theorem}
 The amplitude of a closed connected $D+1$ colored graph, \eqref{eq:ampli} is
   \begin{equation}
    A( {\cal G} ) = N^{D - \frac{2}{(D-1)!} \omega( {\cal G} ) }  \; .
  \end{equation}
\end{theorem}

This is the $1/N$ expansion in random tensor models generalizing in arbitrary dimension
the familiar $1/N$ expansion of matrix models, $A({\cal G} ) = N^{2-2g({\cal G})}$.

\subsection{The leading order graphs}

\begin{figure}[t]
\begin{center}
 \includegraphics[width=7cm]{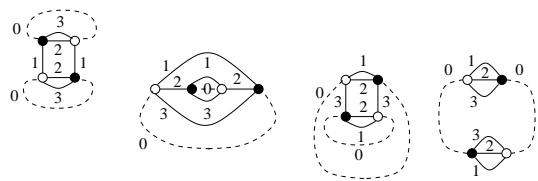}  
\caption{Melons with $p=2$ and $3+1$ colors.}
\label{fig:melonsss}
\end{center}
\end{figure}

All the results presented so far particularize when $D=2$ to the classical matrix models results. In particular $2+1$ colored graphs are ribbon graphs and 
the degree is the genus. At leading order only the $D+1$ colored graphs of degree zero contribute.  
The structure of the $D+1$ colored graphs of degree zero is very different for $D=2$ (matrices) and $D\ge 3$ (tensors). 
The $D+1$ dipole has $2$ vertices and $\frac{D(D-1)}{2}$ faces (one for each couple of colors $ij$) hence degree 
$0$ (figure \ref{fig:degree}(a)). For $D\ge 3$ all the graphs of degree zero
with $2p+2$ vertices can be obtained by inserting two vertices connected by $D$ lines arbitrarily on any line of a $D+1$ colored graph of 
degree zero with $2p$ vertices\cite{Bonzom:2011zz}. This of course does not hold for $D=2$.
We call these graphs {\it melons} (see figure \ref{fig:melonsss}). 

The graph with two vertices and $D+1$ lines represents the coherent identification of two $D$ simplices along their boundary, 
hence it represents a sphere in $D$ dimensions. Two vertices connected by $D$-lines represent a ball in $D$ dimensions. The iterative insertion
of balls into spheres preserves the topology, hence
\begin{theorem}
 For any $D$, the graphs of degree $0$ have spherical topology.
\end{theorem}

As the melonic graphs are generated by an iterative insertion procedure, they can be mapped onto abstract (colored, $D+1$-ary) trees.
The trees are a summable family, hence the melonic graphs are a summable family with 
a finite radius of convergence. The weight of a melonic graph depends on the coupling constants of the model and
tunning to criticality tensor models undergo, like matrix models, a phase transition to continuous infinitely refined random spaces.

\end{document}